\documentclass[%preprint, double-spaced
prb
,twocolumn%
,amssymb, amsmath]{revtex4}
\usepackage[english]{babel}%
\usepackage{bm}%
\usepackage{graphicx}
\usepackage[active]{srcltx}

\newcommand{\mean}[1]{\mathord{\left\langle #1 \right\rangle}}

\newcommand{\llangle}{\mathord{\langle\!\langle}}
\newcommand{\rrangle}{\mathord{\rangle\!\rangle}}

\newcommand{\halb}{\mathord{\frac{1}{2}}}

\newcommand{\fk}{\mathord{\mathbf k}}

\newcommand{\fR}{\mathord{\mathbf R}}
\newcommand{\fS}{\mathord{\mathbf S}}

\newcommand{\zs}{\mathord{z_{\sigma}}}
\newcommand{\comm}[1]{\mathord{\left[ #1 \right]_-}}

\newcommand{\ip}{\mathord{i_{\parallel}}}
\newcommand{\jp}{\mathord{j_{\parallel}}}

\begin{document}

%\title{Importance of the Jahn-Teller effect and direct antiferromagnetic moment coupling for Curie temperatures of manganites}%
\title{Carrier mediated interlayer exchange, ground state phase diagrams and transition temperatures of magnetic thin films} 

\author{M. Stier}
\email{stier@physik.hu-berlin.de}
\author{W. Nolting}%
\affiliation{Festk\"orpertheorie, Institut f\"ur Physik, Humboldt-Universit\"at, 12489 Berlin, Germany}
\date{\today}%

\begin{abstract}
We investigate the influence of the carrier density and other parameters on the interlayer exchange in magnetic thin film systems. The system consists of ferromagnetic and non-magnetic layers where the carriers are allowed to move from layer to layer. For the ferromagnetic layers we use the Kondo-lattice model to describe interactions between itinerant electrons and local moments. The electrons' properties are calculated by a Green's function's equation of motion approach while the magnetization of the local moments is determined by a minimization of the free energy. As results we present magnetic phase diagrams and the interlayer exchange over a broad parameter range. Additionally we can calculate the transition temperatures for different alignments of the ferromagnetic layers' magnetizations.
\end{abstract}

\maketitle

\section{Introduction}

\begin{figure}[tb]
 \includegraphics[width=\linewidth]{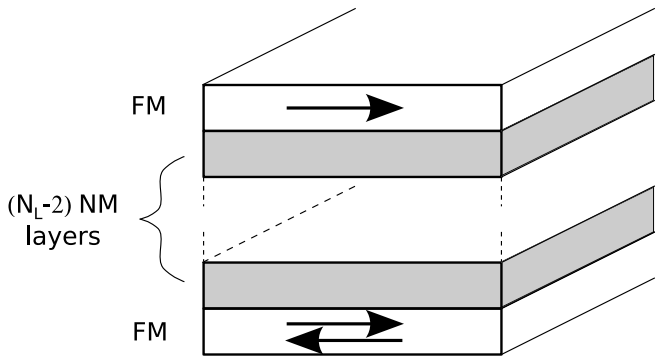}
\caption{\label{figlayer}Schematic picture of the layer system investigated in this work. The system consists of $N_L$ layers with two ferromagnetic (FM) layers at the top and the bottom. They are separated by $N_L- 2$ non-magnetic (NM) layers.}
\end{figure}

Magnetic thin films received much attention in experimental\cite{koch1998,venkatesan2004thin} and theoretical\cite{masisaac1996,bogdanov2001} research, especially after the discovery of a giant magneto resistance (GMR\cite{fert1988,grunberg1989}). One result of this discovery was a massive increase of storage density in hard drives. The phenomenon of the GMR is based on different possible alignments of magnetizations in the single layers. If they are parallel to each other the resistance for a current perpendicular to the layer is considerably smaller than in the antiparallel alignment. Thus it is very interesting to know under which conditions a film material favors to align its layer magnetizations parallel or antiparallel.\\
The way the magnetizations are aligned is defined by the so called interlayer exchange, which can either be ferro- or antiferromagnetic. Even though the interlayer exchange in most materials is assumed to be carrier-mediated\cite{chung2008,akiba1998} it is usually described by a Heisenberg model via an RKKY interaction\cite{bruno1992ruderman,bruno1999,schwieger2004,schwieger2007,kienert2007,sza2009}. This model successfully describes the damped oscillations of the interlayer exchange according to the number of layers\cite{bruno1992ruderman} which essentially influence the type of alignment of the magnetizations.\\
On the other hand the standard RKKY interaction is derived from the free electron gas, which underestimates the interactions in strongly correlated materials. Even modifications of the RKKY\cite{kienert2007}, which incorporate these interactions much more, are based on the Heisenberg model. However, many materials are not best described by a Heisenberg model, but rather, e.g., by a Kondo-lattice/double exchange model. Famous examples are the magnetic $4f$-systems (Gd, Tb,\dots) or the manganites (La$_{1-x}$Ca$_x$MnO$_3$, La$_{1-x}$Sr$_x$MnO$_3$,\dots). Contrary to the Heisenberg model the Kondo-lattice model (KLM) directly describes itinerant carriers aside the local moments. Because the itinerant carriers play an important role for many materials we do not want to use the RKKY method to map the KLM onto a Heisenberg model, since it is unsure how much the original properties of the KLM, and with it the influence of the carriers, are changed within this procedure.\\
The KLM consists of two subsystems - itinerant electrons and local moments. It is difficult to treat both subsystems in one theory on the same level which is one reason for the mapping onto a Heisenberg model in the RKKY method to determine the local moments' properties. In our method, similarly to the RKKY method, we first calculate the electronic properties while we leave the local moments' values (e.g. magnetization $\mean{S^z_i}$) as parameters. But contrary to the RKKY we then calculate the free energy of the system as a function of the magnetization of the local moments\cite{stier3} and find its minimum. Thus we need no mapping on the Heisenberg model.\\
Generally the KLM describes an on-site interaction of local moments $\fS_i$ and an itinerant electron's spin $\boldsymbol \sigma_i$. Depending on the coupling $J$ they prefer to align either parallel ($J>0$) or antiparallel ($J<0$). Due to the hopping of the electrons the whole system can become magnetically ordered. The Hamiltonian of the KLM reads
\begin{align}
 \mathcal H =& \sum_{i,j,\sigma}T_{ij}c^+_{i\sigma}c_{j\sigma}-J\sum_{i}\fS_i\cdot \boldsymbol \sigma_i\label{eqorigklm}\\
		=&\sum_{i,j,\sigma}T_{ij}c^+_{i\sigma}c_{j\sigma}-\frac{J}{2}\sum_{i,\sigma}\left(z_{\sigma}S_i^z n_{i\sigma}+S^{-\sigma}_ic^+_{i\sigma}c_{i,-\sigma }\right)\nonumber\ ,
\end{align}
where $T_{ij}$ is the hopping integral and the $c^{(+)}_{i\sigma}$ are the creators (annihilators) of electrons with spin $\sigma=\pm 1$ at the lattice site $\fR_i$. It is convenient to write the scalar product $\fS_i\cdot \boldsymbol \sigma_i$ explicitly by the use of the Pauli matrices' vector $\hat{\boldsymbol\sigma}$ and $\boldsymbol \sigma_i=\halb\sum_{\sigma\sigma'}c^+_{i\sigma}\hat{\boldsymbol\sigma}_{\sigma\sigma'}c_{i\sigma'}$ to get the second line of (\ref{eqorigklm}). There we additionally introduced $\zs=\delta_{\sigma,+1}-\delta_{\sigma,-1}$, $n_{i\sigma}=c^+_{i\sigma}c_{i\sigma}$ and the spin ladder operator $S^{\sigma}_i=S^x_i+i\zs S^y_i$.\\
One of the main purposes of this work is to calculate the interlayer exchange in dependence on important model parameters. We will only treat "sandwiched" layers in this paper. In particular the system consists of $N_L$ layers where the top and the bottom layers are ferromagnetic (FM), i.e. described by the KLM (cf. Fig. \ref{figlayer}). These two FM layers are separated from each other by $N_L-2$ non-magnetic (NM) layers (spacers). The FM layers can "communicate" with each other by means of the interlayer exchange.\\
Due to the interlayer exchange the magnetizations of both FM layers can align in different ways. We restrict the types of alignment to be either parallel or antiparallel. Other phases like canted ones are neglected in our work. Since the layer system is symmetric the magnetizations should be equal in magnitude, which reduces the amount of numerical calculations significantly.

\section{Theory}

To compare different phases a system can exhibit the key quantity is the internal energy. In the Kondo-lattice model it is simply given by
\begin{align}
\frac{U}{N}=\sum_{\sigma}\int dE\ Ef_-(E)\rho_{\sigma}(E)
\end{align}
where $f_-(E)$ is the Fermi function and $\rho_{\sigma}(E)$ the quasi particle density of states (QDOS). The QDOS can be calculated from the electronic Green's function $G_{\fk\sigma}(E)$ via the condition
\begin{align}
\rho_{\sigma}(E)=-\frac{1}{\pi N}\sum_{\fk}\text{Im}G_{\fk\sigma}(E)\ .
 \end{align}
Thus we need to find an expression for the Green's function, which will in general depend on electronic parameters (e.g. band occupation) and on the spins' properties (e.g. magnetization). First we want to calculate the Green's function only in the electronic subsystem and treat spin expectation values as parameters. In a second step we will show how the optimal spin expectation values can be determined by a minimization of the system's free energy. 

\subsection{\label{secelsys}Electronic subsystem}

We want to find an approximative solution for the electron Green's function
\begin{align}
G_{ij\sigma}(E)=&\llangle c_{i\sigma};c^+_{j\sigma}\rrangle
\end{align}
or its Fourier-transformed equivalent
\begin{align}
G_{\fk\sigma}(E)=&\frac{1}{N}\sum_{i,j}e^{i\fk(\fR_i-\fR_j)}G_{ij\sigma}(E)\ .
\end{align}
To do this we calculate several equations of motion which reads for a general Green's function with the operators $A$ and $B$
\begin{align}
 E\llangle A;B\rrangle=\mean{\left[A,B\right]_{\pm}}+\llangle\comm{A,\mathcal H};B\rrangle\ ,
\end{align}
where $\mean{\dots}$ is the thermodynamic average. The commutator $\comm{A,\mathcal H}$ on the right-hand-side usually leads to higher Green's functions which include more operators and prevent a direct solution of the problem. Thus we have to find a physically reasonable decoupling scheme to trace back these higher functions to those of lower order.

\subsubsection{Film systems}

First we want to discuss the general procedure in a film system\cite{kienert2007}. The Hamiltonian can be written as
\begin{align}
 \mathcal H =& \sum_{\stackrel{i_{\parallel},j_{\parallel},\sigma}{\alpha\beta}}T^{\alpha\beta}_{i_{\parallel}j_{\parallel}}c^+_{i_{\parallel}\alpha\sigma}c_{j_{\parallel}\beta\sigma}+H^{IA}\label{eqlocalfilm}
 \end{align}
with a general interaction term $H^{IA}$. The lattice indices $\ip,\jp$ refer to lattice sites $\fR_{\ip,\jp}$ within a film plane while $\alpha,\beta$ name a distinct layer. Different layer positions are described by a primitive translation $\mathbf r^{\alpha}$ perpendicular to the film planes ($\fR_{\ip}^{\alpha}=\fR_{\ip}+\mathbf r^{\alpha}$). A movement of the electrons is incorporated by a hopping term $T^{\alpha\beta}_{i_{\parallel}j_{\parallel}}$ where the electron can move within the film from $\fR_{\ip}^{\alpha}$ to $\fR_{\jp}^{\alpha}$ or between two layers $\alpha,\beta$ ($\fR_{\ip}^{\alpha}\rightarrow\fR_{\ip}^{\beta}=\fR^{\alpha}_{\ip}+\mathbf r^{\alpha\beta}$). To simplify the calculations we restrict the movement to next-neighbor hopping  in a simple cubic lattice
\begin{align}
 T_{i_{\parallel}j_{\parallel}}^{\alpha\beta}=&\frac{W^{2D}}{8}\delta^{\alpha\alpha}_{i_{\parallel},j_{\parallel}\pm 1}+\frac{W^{2D}}{8}\delta^{\alpha,\beta\pm 1}_{i_{\parallel}i_{\parallel}}
\end{align}
with $\delta_{\ip\jp}^{\alpha\beta}=\delta_{\ip\jp}\delta_{\alpha\beta}$, the free band width of a single film $W^{2D}$ and the resulting dispersion
\begin{align}
\epsilon^{\alpha\beta}(\fk_{\parallel})=&\frac{1}{N_{\parallel}}\sum_{\mean{i_{\parallel},j_{\parallel}}}T^{\alpha\beta}_{i_{\parallel}j_{\parallel}}e^{i\fk_{\parallel}(\fR_{i_{\parallel}}-\fR_{j_{\parallel}})}\ .\label{eqdisp}
\end{align}
The Fourier transformation is restricted to the layer with $N_{\parallel}$ sites and the 2-dimensional wave vector $\fk_{\parallel}$. We can set up the equation of motion for the electron Green's function
\begin{align}
G^{\alpha\beta}_{\ip\jp\sigma}(E)&=\llangle c_{\ip\alpha\sigma};c^+_{\jp\beta\sigma}\rrangle\ ,\\
 \sum_{l_{\parallel},\gamma}\Big(E\delta^{\alpha\gamma}_{l_{\parallel}\ip}&-T_{l_{\parallel}\jp}^{\gamma\beta}\Big)G_{l_{\parallel}\jp\sigma}^{\gamma\beta}(E)=\label{eqgenself}\\
&=\delta_{\ip\jp}^{\alpha\beta}+\llangle\comm{c_{\ip\alpha\sigma},H^{IA}};c_{\jp\beta\sigma}\rrangle\nonumber\\
&=\delta_{\ip\jp}^{\alpha\beta}+M_{\alpha\sigma}(E)G_{\ip\jp\sigma}^{\alpha\alpha}(E)\ .\nonumber
\end{align}
In the last step we introduced a self-energy $M_{\alpha\sigma}(E)$ which leads to a formal solution of the equation. Generally the self-energy can be non-local, $M_{\ip\jp\sigma}^{\alpha\beta}(E)$, but only local ones will occur in this work. With a Fourier-transformation as in (\ref{eqdisp}) we find the Green's function by an inversion of an ($N_L\times N_L$)-matrix $A_{\sigma}$
\begin{align}
 G_{\fk_{\parallel}\sigma}^{\alpha\beta}(E)&=\left(A_{\sigma}^{-1}\right)_{\alpha\beta}\label{eqgmatrix}\\
A_{\sigma}&=
\left(
\begin{array}{cccc}
 g^{-1}_{\fk_{\parallel}1\sigma}(E)& \epsilon_{\perp} & 0 &\dots\\ 
\epsilon^*_{\perp}  &  g^{-1}_{\fk_{\parallel}2\sigma}(E)&  \epsilon_{\perp} & 0 \\ 
0 &   \epsilon_{\perp}^*& g^{-1}_{\fk_{\parallel}3\sigma}(E)& \dots  \\ 
\dots & 0 & \dots & \dots                                   \end{array}\right)\nonumber
\end{align}
with
\begin{align}
 g^{-1}_{\fk_{\parallel}\alpha\sigma}(E)=E-\epsilon_{\parallel}(\fk_{\parallel})-M_{\alpha\sigma}(E)\ ,\\
\epsilon_{\parallel}(\fk_{\parallel})=\epsilon^{\alpha\alpha}(\fk_{\parallel}),\ \epsilon_{\perp}=\epsilon^{\alpha,\beta\neq\alpha}\ .
\end{align}
When we find an expression for the self-energy the problem is solved.

\subsubsection{Kondo lattice system}

To find a solution for the self-energy, we explicitly use the Kondo-lattice model adapted to film geometry now:
\begin{align}
 \mathcal H =& \sum_{\stackrel{i_{\parallel},j_{\parallel},\sigma}{\alpha\beta}}T^{\alpha\beta}_{i_{\parallel}j_{\parallel}}c^+_{i_{\parallel}\alpha\sigma}c_{j_{\parallel}\beta\sigma}-\\
&-\frac{J}{2}\sum_{i_{\parallel},\alpha,\sigma}\left(z_{\sigma}S_{i_{\parallel}\alpha}^z n_{i_{\parallel}\alpha\sigma}+S^{-\sigma}_{i_{\parallel}\alpha}c^+_{i_{\parallel}\alpha\sigma}c_{i_{\parallel}\alpha,-\sigma}\right)\nonumber\ .
 \end{align}
Up to now no general solution for the KLM is known and that is why we use a moment conserving decoupling approach (MCDA\cite{mcda1997}) to get the self-energy $M_{\alpha\sigma}(E)$. This non-perturbative approach is based on equations of motions up to the second order and several decoupling schemes. It has been used successfully various times to describe different aspects of the KLM\cite{carlos2002,mrkky2003,kienert2007,stier3}. We want to sketch the general procedure and refer to Ref. \cite{mcda1997} for details. Starting point is the equation of motion of the electron Green's function
\begin{align}
  \sum_{l_{\parallel},\gamma}&\left(E\delta^{\alpha\gamma}_{l_{\parallel}\ip}-T_{l_{\parallel}\jp}^{\gamma\beta}\right)G_{l_{\parallel}\jp\sigma}^{\gamma\beta}(E)=\\
&=\delta_{\ip\jp}^{\alpha\beta}\underbrace{-\frac{J}{2}\left(\zs I^{\alpha\alpha\beta}_{\ip\ip\jp\sigma}(E)+F^{\alpha\alpha\beta}_{\ip\ip\jp\sigma}(E)\right)}_{=M_{\alpha\sigma}(E)G^{\alpha\beta}_{\ip\jp\sigma}(E)}\label{eqfirsteom}
\end{align}
with
\begin{align}
 I^{\alpha\gamma\beta}_{\ip m_{\parallel} \jp\sigma}(E)&=\llangle S^z_{\ip\alpha}c_{m_{\parallel}\gamma\sigma};c^+_{\jp\beta\sigma}\rrangle\\
F^{\alpha\gamma\beta}_{\ip m_{\parallel} \jp\sigma}(E)&=\llangle S^{-\sigma}_{\ip\alpha}c_{m_{\parallel} \gamma,-\sigma};c^+_{\jp\beta\sigma}\rrangle\ .
\end{align}
Two new higher Green's functions appear; the Ising function $I^{\alpha\gamma\beta}_{\ip m_{\parallel} \jp\sigma}(E)$ and the spin-flip function $F^{\alpha\gamma\beta}_{\ip m_{\parallel} \jp\sigma}(E)$. Due to the general expression of $G^{\alpha\beta}_{\ip\jp\sigma}(E)$ via a self-energy as in (\ref{eqgenself}) they have to be connected to $M_{\alpha\sigma}(E)$ as indicated in (\ref{eqfirsteom}). Thus when we find appropriate expressions for both functions we get the self-energy, too. To do this we set up the equations of motion for both functions:
\begin{align}
 \sum_{l_{\parallel},\mu}&\Big(E\delta^{\mu\gamma}_{l_{\parallel}m_{\parallel}}-T_{l_{\parallel}\jp}^{\mu\beta}\Big)I_{\ip l_{\parallel}\jp\sigma}^{\alpha\mu\beta}(E)=\label{eqeomi}\\
&=\mean{S_{\ip\alpha}^z}\delta_{m_{\parallel}\jp}^{\gamma\beta}+\llangle\comm{S_{\ip\alpha}^zc_{m_{\parallel}\gamma\sigma},H^{IA}};c_{j\beta\sigma}\rrangle\nonumber\\
 \sum_{l_{\parallel},\mu}&\Big(E\delta^{\mu\gamma}_{l_{\parallel}m_{\parallel}}-T_{l_{\parallel}\jp}^{\mu\beta}\Big)F_{\ip l_{\parallel}\jp\sigma}^{\alpha\mu\beta}(E)=\label{eqeomf}\\
&=\llangle\comm{S_{\ip\alpha}^{-\sigma}c_{m_{\parallel}\gamma,-\sigma},H^{IA}};c_{j\beta\sigma}\rrangle\nonumber
\end{align}
The commutators on the right-hand-side prevent a direct solution. We now differentiate between diagonal ($\ip=m_{\parallel}$ and $\alpha=\gamma$) and non-diagonal terms ($\ip\neq m_{\parallel}$ or $\alpha\neq \gamma$). For the non-diagonal ones we approximate
\begin{align}
 \llangle S^z_{\ip\alpha}\comm{c_{m_{\parallel}\gamma\sigma},H^{IA}}&;c^+_{\jp\beta\sigma}\rrangle\approx\label{eqnond1}\\
&\approx M_{\gamma\sigma}(E)I^{\alpha\gamma\beta}_{\ip m_{\parallel} \jp\sigma}(E)\nonumber\\
 \llangle S^{-\sigma}_{\ip\alpha}\comm{c_{m_{\parallel}\gamma,-\sigma},H^{IA}}&;c^+_{\jp\beta\sigma}\rrangle\approx\label{eqnond2}\\
&\approx M_{\gamma,-\sigma}(E)F^{\alpha\gamma\beta}_{\ip m_{\parallel} \jp\sigma}(E)\nonumber
\end{align}
and
\begin{align}
 \llangle \comm{S^z_{\ip\alpha},H^{IA}}c_{m_{\parallel}\gamma\sigma};c^+_{\jp\beta\sigma}\rrangle&\approx 0\label{eqnond3}\\
 \llangle \comm{S^{-\sigma}_{\ip\alpha},H^{IA}}c_{m_{\parallel}\gamma,-\sigma};c^+_{\jp\beta\sigma}\rrangle&=0\label{eqnond4}\ .
\end{align}
In (\ref{eqnond1}), (\ref{eqnond2}) we used the formal equivalence
\begin{align}
 \comm{c_{m_{\parallel}\gamma,\pm\sigma},H^{IA}}\rightarrow M_{\gamma,\pm\sigma}(E)c_{m_{\parallel}\gamma\pm\sigma}
\end{align}
due to the Dyson equation, while the Green's functions in (\ref{eqnond3}), (\ref{eqnond4}) can be connected to magnon energies, which are much lower than the electronic energies\cite{mcda1997}. This holds for all spin quantum numbers $S$.\\
Due to higher correlations the same procedure cannot be done for the diagonal terms. But in this case we can calculate the commutators in  (\ref{eqeomi}) and (\ref{eqeomf}) explicitly. As usual new higher Green's functions appear
\begin{align}
 F^{(1)}_{iiij\sigma}(E) &= \llangle S_i^{\bar\sigma}S_i^z c_{i\bar\sigma};c_{j\sigma}^+\rrangle\label{eqF1}\\
 F^{(2)}_{iiij\sigma}(E) &= \llangle S_i^{\bar\sigma}S_i^{\sigma} c_{i\sigma};c_{j\sigma}^+\rrangle\label{eqF2}\\
 F^{(3)}_{iiiij\sigma}(E) &= \llangle S_i^{\bar\sigma}n_{i\sigma}c_{i\bar\sigma};c_{j\sigma}^+\rrangle\label{eqF3}\\
 F^{(4)}_{iiiij\sigma}(E) &= \llangle S_i^z n_{i\bar\sigma}c_{i\sigma};c_{j\sigma}^+\rrangle \label{eqF4}
\end{align}
where we suppressed the layer indices for readability ($i\equiv\ip\alpha,\ j\equiv\jp\beta$). The first two functions can be written in the limiting cases of low spin $S=\halb$ or ferromagnetic saturation $\mean{S^z}=S$ as linear combinations of the electron Green's function, the Ising function and/or the spin-flip function. Thus we assume that a linear combination is also possible aside these limiting cases and choose
\begin{align}
 F^{(1)}_{iiij\sigma}(E) 
&=a^{(1)}_{\alpha\sigma}G^{\alpha\beta}_{\ip\jp\sigma}(E)+b^{(1)}_{\alpha\sigma}F^{\alpha\alpha\beta}_{\ip\ip\jp\sigma}(E)\label{eqlincomb1}\\
 F^{(2)}_{iiij\sigma}(E) 
&=a^{(2)}_{\alpha\sigma}G^{\alpha\beta}_{\ip\jp\sigma}(E)+b^{(2)}_{\alpha\sigma}I^{\alpha\alpha\beta}_{\ip\ip\jp\sigma}(E)\ .
\end{align}
The coefficients can be explicitly determined by the condition that the moments of these functions have to be conserved. Contrary to Ref. \cite{mcda1997} we set
\begin{align}
 F^{(3)}_{iiiij\sigma}(E) =& \llangle S_i^{\bar\sigma}n_{i\sigma}c_{i\bar\sigma};c_{j\sigma}^+\rrangle&\approx 0\\
 F^{(4)}_{iiiij\sigma}(E) =&\llangle S_i^z n_{i\bar\sigma}c_{i\sigma};c_{j\sigma}^+\rrangle&\approx 0
\end{align}
which effectively means a neglection of double occupation effects, for the sake of simplicity. Recent results\cite{stier4} show that double occupation effects play a minor role at high spins. On the other hand even \emph{with} inclusion of double occupancies, which is essential for lower spins, results (phase diagrams, transition temperatures) hardly depend on the spin quantum number $S$ but on the coupling strength $JS$ for $J>0$. Thus we will only use $S\gg \halb$ and $J>0$ later on, which allows the neglection of double occupancies while the dependence on $JS$ holds for all spins.\\
Due to the simplifications all terms are traced back onto four unknown functions: $G^{\alpha\beta}_{\ip\jp\sigma}$, $ F^{\alpha\alpha\beta}_{\ip\ip\jp\sigma}$, $ I^{\alpha\alpha\beta}_{\ip\ip\jp\sigma}$ and $M_{\alpha\sigma}$. The equations of motion of the spin-flip and Ising function reduce to analytical expression which have the solutions
\begin{align}
 X^{\alpha\alpha\beta}_{\ip\ip\jp\sigma}(E)= \bar X_{\alpha\sigma}(E) G^{\alpha\beta}_{\ip\jp\sigma}(E),\quad X=I,F\ .
\end{align}
Both are proportional to the electron Green's function $G^{\alpha\beta}_{\ip\jp\sigma}(E)$. The prefactors $\bar X_{\alpha\sigma}(E)$ only contain local terms, like expectation values, the local self-energy $M_{\alpha\sigma}(E)$ and the local Green's function
\begin{align}
 G_{\alpha\sigma}(E)&=\frac{1}{N_{\parallel}}\sum_{\fk_{\parallel}}G^{\alpha\alpha}_{\fk_{\parallel}\sigma}(E)\nonumber\ .
\end{align}
With (\ref{eqfirsteom}) we can write the self-energy
\begin{align}
 M_{\alpha\sigma}(E)=-\frac{J}{2}\left(\zs \bar I_{\alpha\sigma}(E)+\bar F_{\alpha\sigma}(E)\right)
\end{align}
which is then indeed local, too. In fact it was not necessary to assume a local self-energy in (\ref{eqfirsteom}) but it simplifies the derivation of $M_{\alpha\sigma}(E)$. The reason for the locality of the self-energy is only due to the neglection of magnon energies\cite{mcda1997}.\\
Especially due to the dependence of $M_{\alpha\sigma}(E)$ on the local electronic Green's function 
\begin{align}
 M_{\alpha\sigma}(E)& = M_{\alpha\sigma}\{ G_{\alpha,\pm\sigma}(E)\}\ .
\end{align}
the self-energy is through (\ref{eqgmatrix}) connected to the Green's functions of all other layers. Since we get all Green's functions $G^{\alpha\alpha}_{\fk_{\parallel}\sigma}(E)$ from (\ref{eqgmatrix}) the whole system of equations is closed. All implicitly included, e.g. in $a^{(1)}_{\alpha\sigma}$ in (\ref{eqlincomb1}), electronic expectation values ($\mean{n_{\alpha\sigma}}$, $\mean{S^z_{\alpha}n_{\sigma\alpha}}$, \dots) are obtainable by the spectral theorem from the involved Green's functions. The spins' expectation values ($\mean{(S^z)^2}(\varphi)$, $\mean{(S^z)^3}(\varphi)$, \dots) can be derived from the magnon density $\varphi$ which we can get from a given magnetization\cite{mcda1997,callen} $\mean{S^z}\equiv M(\varphi)\Leftrightarrow \varphi(M)$, which we see as a parameter.\\
With the calculation of the Green's function we also get the important quantity of the internal energy
\begin{align}
 \frac{U}{N}=-\frac{1}{\pi N_{\parallel}N_L}\sum_{\fk_{\parallel},\alpha,\sigma}\int dE\ E f_-(E)\text{Im}G_{\fk_{\parallel}\sigma}^{\alpha\alpha}(E)
\end{align}
which can be used to compare the energies of different phases.\\
In the special systems we treat in this work (cf. Fig. \ref{figlayer}) only the layers at the top and the bottom have a non-vanishing (direct) self-energy ($M_{\alpha\sigma}(E)\equiv 0, \forall\alpha\neq1,N_L$), but also the NM layers are influenced by interactions within the FM layers due to a hybridization induced by the hopping $\epsilon_{\perp}$ in (\ref{eqgmatrix}). Additionally the FM and NM layers can differ in their centers of gravity $T_{\ip\ip}^{\alpha\alpha}$ of the free bands. Throughout the paper we use
\begin{align}
  T_{\ip\ip}^{\alpha\alpha}\stackrel{\alpha=1,N_L}{\equiv}T_0^{\text{FM}}=0 
\end{align}
while
\begin{align}
 T_{\ip\ip}^{\alpha\alpha}\stackrel{\alpha\neq 1,N_L}{\equiv}T_0^{\text{NM}} 
\end{align}
may be unequal zero.

\subsection{Local moments}

In the previous section we derived an expression for the electronic Green's function and thereby also for the internal energy. One drawback of the method is that we have to see the magnetization of the local moments as a parameter. Thus we cannot determine $M\equiv \mean{S^z}$ from the equations itself. This can be solved if we can get the free energy $F(M,T)$ for many values of $M$ and decide which magnetization leads to the lowest energy. 

\subsubsection{Free energy}

The free energy is given by
\begin{align}
 F(T,M)=U(M,T)-TS(M,T)
\end{align}
with the entropy $S(M,T)$. We can use $S_M(T)=-\frac{\partial F_M(T)}{\partial T}$ and get
\begin{align}
 F_M(T)=U_M(T)+T\frac{\partial F_M(T)}{\partial T}
\end{align}
which can be transformed via the product rule and
\begin{align}
 \lim_{T\rightarrow 0}\left(\frac{1}{T}(F_M(T)-F_M(T=0))\right)=&\left.\frac{\partial F_M(T)}{\partial T}\right|_{T=0}\\
=&-S_M(T=0)\nonumber
\end{align}
to
\begin{align}
 F_M(T)=&U_M(0)-TS_M(0)-\label{eqfm}\\
&-\underbrace{\int_0^T dT' \frac{U_M(T')-U_M(0)}{(T')^2}}_{\equiv I_M(T)}\nonumber\ .
\end{align}
Since we can get the internal energy from the method of Sec. \ref{secelsys} the free energy can be calculated when we find an expression for the entropy at $T=0$. When we assume this can be done we can calculate the free energy for many values of $M$ for a distinct temperature $T$. Thus we get an ensemble of energies $\{F_M(T)\}$ which leads to a magnetization dependent free energy
\begin{align}
 \{F_M(T)\}\rightarrow F_T(M)\ .
\end{align}
When we find its minimum according to $M$
\begin{align}
 \left.\frac{\partial F_T(M')}{\partial M'}\right|_{M}=0
\end{align}
 we get the optimal magnetization of the system.\\
Normally one would use the system's total magnetization  $M^{\text{tot}}=M+\sigma^z(M)$ of local moment and electron magnetization $\sigma^z$ for the minimization of the free energy, but since the total magnetization in our case is a function of $M$ it is sufficient to use $M$ as the order parameter.

\subsubsection{\label{secentropy}Entropy}

The aim of this section is to find an approximate expression for the entropy at $T=0$. Generally the entropy can be written as
\begin{align}
 S_0(M)=k_B\ln \Gamma_M\label{eqdefentr}
\end{align}
where $\Gamma_M$ is the number of states which contribute to a magnetization $M$ and $k_B$ is the Boltzmann constant. Note that the expression (\ref{eqdefentr}) is \emph{not} the ground state entropy! Due to the third law of thermodynamics the ground state entropy would be a constant according to the system parameters. But the system is not in the ground state for most magnetizations, but only for one optimal $M^{\text{opt}}$. Thus the ground state entropy is $S_0(M^{\text{opt}})$ while the complete \emph{magnetization dependent} entropy $S_0(M)$ is not visible in reality but plays the role of an auxiliary quantity to find the optimal value $M^{\text{opt}}$.\\
The whole systems consists of itinerant electrons and local moments which both affect the entropy. First we make the ansatz that the total number of states can be written as a product of the subsystems' states $\Gamma_M=\Gamma_M^{\text{loc}}\Gamma_M^{\text{el}}$, which means that the number of states of one subsystem is not affected by a special state of the other one. This ansatz can be motivated by the dynamics of the subsystems, which differ in several orders of magnitude between the fast electrons and the slow moments. Thus the total entropy is given by a sum of the single ones:
\begin{align}
 S_0(M)=S_0^{\text{loc}}(M)+S_0^{\text{el}}(M)\ .
\end{align}
For the electron entropy we use the expression for a Fermi gas
\begin{align}
 S_0^{\text{el}}(M)=-k_B\sum_{\fk_{\parallel},\alpha,\sigma}&\left[(1-\mean{n_{\fk_{\parallel}\alpha\sigma}})\ln(1-\mean{n_{\fk_{\parallel}\alpha\sigma}})+\right.\nonumber\\
&+\left.\mean{n_{\fk_{\parallel}\alpha\sigma}}\ln\mean{n_{\fk_{\parallel}\alpha\sigma}}\right]
\end{align}
which seems plausible since the electrons see more or less a static potential of the moments. The mean values $\mean{n_{\fk_{\parallel}\alpha\sigma}}$ are, however, calculated by the full theory of Sec. \ref{secelsys} at a fixed $M$!\\
Each local moment can have a spin projection $m_i^z=-S,-S+1,\dots,S$ while the complete system can be described by a configuration $\{m_i^z\}$. We now assume that all configurations with the same magnetization $\{m_i^z|N^{-1}\sum_i m_i^z=M\}$ have the same energy while configurations with different magnetizations have different energies. This can be for example, but not exclusively, the case if the electrons act as an effective field $B^{\text{eff}}$ on the local moments (ideal paramagnet). The size or sign of the field plays no role as long as $B^{\text{eff}}\neq 0$. It means that a special magnetization is equivalent to a distinct energy and we can count the number of states as in the microcanonical ensemble
\begin{align}
 \Gamma(E)=&\sum_l^{E<E_l<E+\Delta E}1\\
\Leftrightarrow \Gamma^{\text{loc}}(M)=&\sum_l^{\{m_i|M=N^{-1}\sum_i m_i \}_l} 1\ .
\end{align}
Thus the number of states is equal to the number of configurations which leads to a magnetization $M$. These assumptions can only be made because we want to find the entropy at $T=0$. At larger temperatures more energy levels contribute to a thermodynamical mean value like the magnetization due to energy/temperature dependent factors like $\sim e^{(k_B T)^{-1}E_l}$.\\
We find that, for example, the entropy for spins with $S=\halb$ is given as
\begin{align}
 \frac{S_0(M)}{N}\stackrel{N\rightarrow\infty}{=}-k_B&\left[(\halb-M)\ln(\halb-M)\right.+\nonumber\\
&+\left.(\halb+M)\ln(\halb+M)\right]\ .
\end{align}
Note that in the film systems treated in this work not every lattice site has to be occupied by a local moment which reduces the local moment entropy. The complete lattice consists of $N=N_LN_{\parallel}$ lattice site ($N_{\parallel}$ is the number of sites of \emph{one} layer). Due to the independence of local moments the total entropy can be expressed by a sum of layer entropies $S_0^{\alpha,\text{loc}}(M)$
\begin{align}
 \frac{S_0^{\text{loc}}(M)}{N}=\frac{1}{N_L}\sum_{\alpha=1}^{N_L}\frac{S_0^{\alpha,\text{loc}}(M)}{N_{\parallel}}
\end{align}
where the entropies of the NM layers are zero.\\
The approximations used in this section for the entropy are reasonable for the Kondo lattice model, but cannot be assigned to other systems, e.g. the Heisenberg model, without further considerations of their special properties.

\section{Results}

\begin{figure}[tb]
 \includegraphics[width=\linewidth]{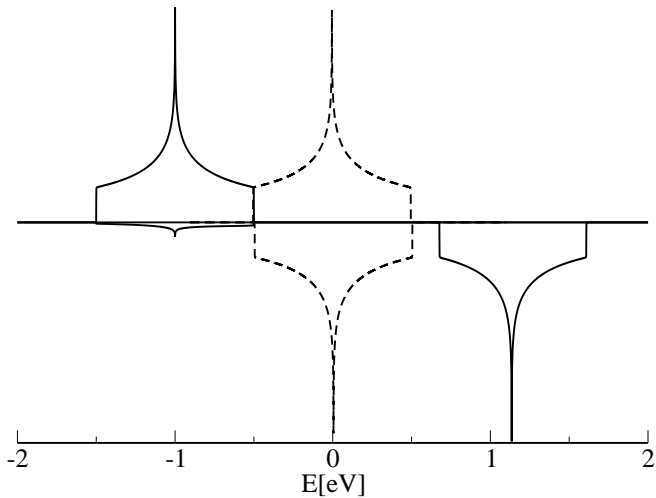}
\caption{\label{figdosuncoupl}Spin up/down quasi particle density of states of independent layers for an FM (\emph{solid line}, $M=S$) and an NM layer (\emph{dashed line}). Parameters: $S=7.5$, $JS=2$eV, $n=0.5$, $W^{2D}=1$eV, $T_0^{\text{NM}}=0$ }
\end{figure}
We want to discuss the existence of magnetic configurations. At vanishing temperature $T=0$ we only have to compare the internal energies which can be directly calculated from the electronic Green's function. At higher temperatures $T>0$ there exists an interplay between a reduction of internal energy and an increase of entropy. Therefore we have to use the free energy as the decisive thermodynamic potential in this case.

\subsection{\label{sect0}Phase diagrams at T=0}

\begin{figure*}[tb]
 \includegraphics[width=1\linewidth]{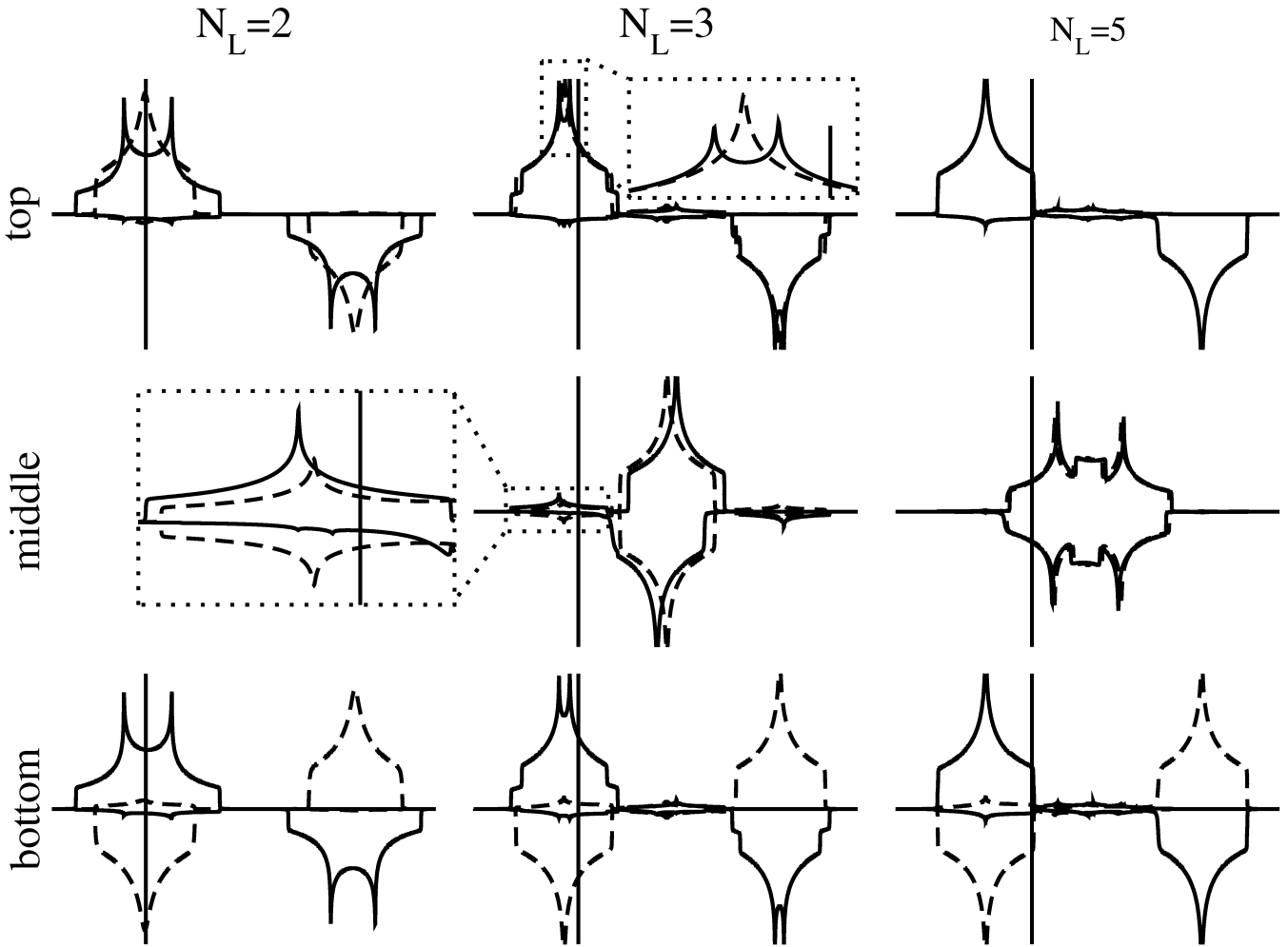}
\caption{\label{figdoslayer}Quasi particle density of states of the top (FM), middle (NM) and bottom (FM) layer for parallel (\emph{solid line}) and antiparallel configuration (\emph{dashed line}). Differences to uncoupled layers occur due to a hybridization between the layers (cf. the magnified QDOS of the NM layer for $N_L=3$). At large couplings (shown here) the FM layers are energetically far away from the NM layers and the hybridization is very small. A special case is $N_L=2$ where no NM layer exists and the FM layers hybridize directly. Vertical lines mark the chemical potential. Parameters: $S=7.5$, $JS=2$eV, $n=0.5$, $W^{2D}=1$eV, $T_0^{\text{NM}}=0$ }
\end{figure*}
Before we treat coupled layer systems it is instructive to mention some properties of independent layers, since important characteristics are valid for the coupled ones, too. First of all it is a basic question whether or not a layer described by a Kondo-lattice model is actually ferromagnetic for a given parameter constellation. We want to concentrate on two parameters: the coupling $J$ (or better $JS$) and the band occupation $n$. In the Kondo lattice model (with neglection of double occupancies) the electron band splits into two subbands which are for high spins at energies $E\approx \pm \halb JS$ (cf. Fig. \ref{figdosuncoupl}). It is known for positive couplings $J$ that the formation of ferromagnetic order needs a distinct coupling strength $JS$. For low couplings $JS$ ferromagnetism only occurs at low band fillings $n$. The higher the coupling the higher the band filling can be before the ferromagnetic order breaks down. A maximum point is reached at half-filling $n=1$ where ferromagnetism never exists\cite{chatto2001,carlos2002,stier2,peters2007,kalpatru2009,henning2009,stier3}. Thus before we can compare different configurations of coupled films we have to find out if the \emph{intrinsic} ferromagnetic order is stable. This is done in this work by a comparison of the energies of the ferromagnetic ($M=S$) and the paramagnetic ($M=0$) solutions. Normally at $n\approx 1$ antiferromagnetic phases occur which we neglect in this work.\\
In uncoupled FM films neither parallel nor antiparallel alignment of magnetizations is favored, since the total QDOSs (i.e. spin-up plus spin-down) and with it the internal energies are the same. Things change in coupled systems. The layers interact with each other via a hopping $\epsilon_{\perp}(\fk_{\parallel})$ in (\ref{eqgmatrix}) perpendicular to the film plane. This leads to a hybridization between the single layers' densities of states, resulting in a transfer and a repulsion of states to other energies (cf. Fig. \ref{figdoslayer}). The strength of the hybridization, i.e. the number of transferred states and the amount of repulsion, depends on the original (energetic) distance of the states of the uncoupled films - the closer the states the stronger the hybridization. \\
Especially parallel and antiparallel configurations differ in coupled layer systems, because the hybridization acts only between the QDOSs of the same spin. For the parallel alignment the QDOSs per spin direction of the two films are energetically close to each other, while they are far away for the antiparallel configuration. That is why the QDOS of the system with parallel alignment is considerably changed due to the hybridization. This can be directly seen for the two layer case  in Fig. \ref{figdoslayer}.\\
\begin{figure*}[tb]
 \includegraphics[width=1\linewidth]{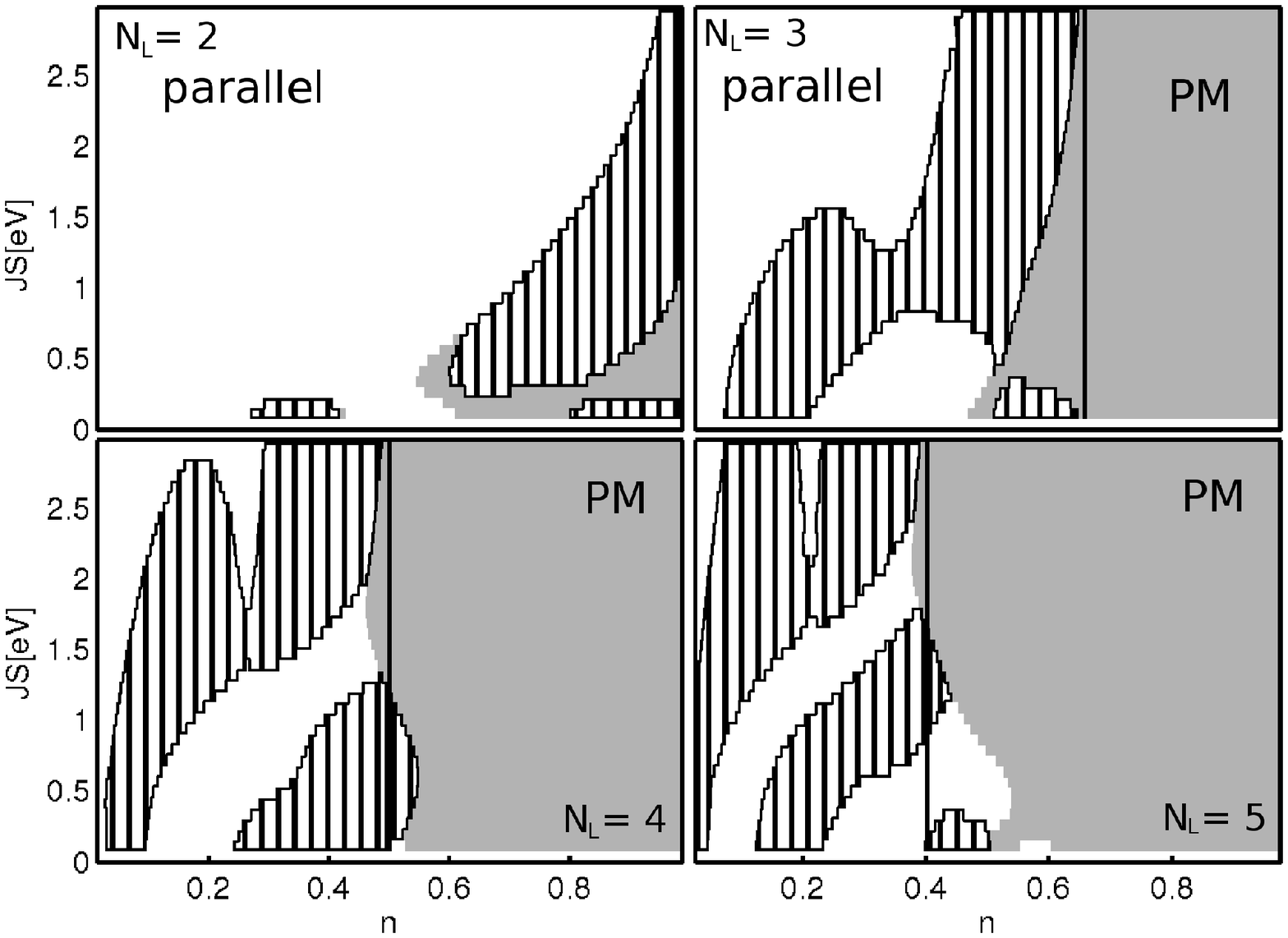}
\caption{\label{figpdlayer}Phase diagram of the orientation of the magnetizations of the ferromagnetic layers with $N_L=2,3,4,5$. The magnetizations are either parallel (\emph{white}) or antiparallel (\emph{striped}). Additionally, areas where the layers themselves are not magnetically ordered are shown (\emph{gray}). Vertical lines mark the critical band occupation where the FM layers are completely filled at large couplings ($n_C=2/N_L$). Parameters: $S=7.5$, $W^{2D}=1$eV, $T_0^{\text{NM}}=0$ }
\end{figure*}
Changes in the QDOS are also changes in the internal energy. So one can expect a strong influence of the hybridization on the phase diagram. We consider three types of phases: 
\begin{itemize}
 \item both FM layers are ferromagnetically saturated ($M=S$) and both magnetizations are aligned parallel to each other
\item as before but with antiparallel alignment
\item the FM layers are not magnetically ordered ($M=0$, paramagnetic\footnote{Even though the layers are not ferromagnetic any more they are still called "FM" layers because they are in principle able to be ferromagnetic in contrast to the NM layers.})
\end{itemize}
Indeed all three phases occur in the phase diagram (cf. Fig. \ref{figpdlayer}). Let us first discuss the $N_L=2$ system, where the total system consists of two FM layers only. As in a bulk system\cite{yunoki1998,chatto2001,fishman2006,henning2009} parallel alignment (corresponding to the ferromagnetic phase in the bulk) exists always at low doping $n$ and increases its density range for strong couplings $J$. At $n\lesssim 1$ the antiparallel alignment is preferred which can be compared to layered antiferromagnetic phases in the bulk\cite{henning2009}. Thus the parameter dependencies on $J$ and $n$ stay qualitatively the same as in the bulk system. As said before not all phases which have been found in bulk systems are considered in this work for simplicity.\\
When we go to $N_L>2$ the FM layers are separated by $N_L-2$ NM layers. Two new effects occur in this case. First it can be seen in Fig. \ref{figdoslayer} that for strong couplings parts of the QDOS of the FM layers are far below the QDOS of the NM layers. This means that the FM lattice sites are occupied by electrons at first which leads to a charge transfer from the NM to the FM layers. Thus the total electron density $n$ normally differs from the density $n^{\text{FM}}$ within the FM layers. The total density in the system is 
\begin{align}
n=N^{\text{el}}/N
\end{align}
where $N^{\text{el}}$ is the absolute number of electrons and $N$ is the total number of lattice sites. At strong couplings only the FM sites are occupied leading to a density in the FM layers 
\begin{align}
 n^{\text{FM}}=N^{\text{el}}/N^{\text{FM}}
\end{align}
with the number of lattice sites of the FM layers $N^{\text{FM}}$. In our sandwiched layer systems only two FM layers are present in an $N_L$ layer system which means that $N^{\text{FM}}/N=2/N_L$. This means that the FM layers are half-filled
\begin{align}
 n^{\text{FM}}=1 \Leftrightarrow n=2/N_L\equiv n_C\ .
\end{align}
 Indeed this band occupation $n_C$ marks the transition to the paramagnetic region at large couplings. A similar effect has been found for the diluted Kondo-lattice model\cite{dilklm}.\\
The second effect is due to the hybridization. When more layers are present in a system the hybridization leads to more complex shapes of the QDOS (cf. for example the middle layer in Fig. \ref{figdoslayer}). This manifests itself by a more complex structure of the phase diagram at low couplings or $n<n_C$. Areas of parallel and antiparallel configurations of systems with larger $N_L$ alternate more often than in systems with small $N_L$. Especially the increased number of peaks in the QDOS, compared to independent layers, influences the internal energy and with it the preferred configuration.\\
\begin{figure}[tb]
 \includegraphics[width=\linewidth]{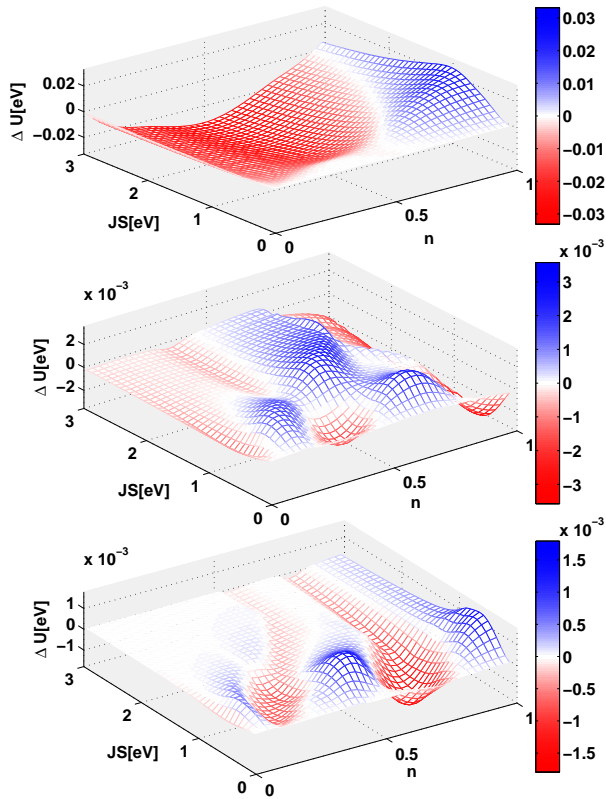}
\caption{\label{figexITNM0}(color online) Interlayer exchange $\Delta U = U^{\uparrow\uparrow}-U^{\uparrow\downarrow}$ for $N_L=2$ (\emph{top}), $3$ (\emph{middle}) and $4$ (\emph{bottom}) layers in dependence of carrier density $n$ and coupling $JS$. Negative values (\emph{red}) mean ferromagnetic and positive (\emph{blue}) antiferromagnetic exchange. Parameters: $S=7.5$, $W^{2D}=1$eV, $M=S$, $T_0^{\text{NM}}=0$ }
\end{figure}
But there is another question. Even though we get a larger complexity of the QDOS with increasing $N_L$ the absolute differences between the QDOSs for the two alignment types get smaller (cf. Fig. \ref{figdoslayer}, for $N_L=5$ the total QDOSs (spin-up plus spin-down) of the two alignments are almost equal). From the phase diagram we do not know how large the energetic differences between these alignments are. Thus we define the interlayer exchange
\begin{align}
 \Delta U =U^{\uparrow\uparrow}-U^{\uparrow\downarrow}\label{eqdefie}
\end{align}
as the difference between the two energies of both configurations. In this definition a negative $\Delta U$ means a ferromagnetic interlayer exchange. We indeed see the expected dependencies in Fig. \ref{figexITNM0}. Even though an intrinsic magnetic order of a single FM layer is enhanced\cite{henning2009} by a large $J$, the interlayer exchange is heavily damped. This is due to the reduction of hybridization effects which are responsible for the intralayer exchange. For high $N_L$ the damping is larger because an electron has to cross more NM layer to get from one FM layer to the other. The interlayer exchange has its maximum (absolute) value at intermediate couplings where hybridization effects are still relatively strong and the intrinsic ferromagnetism is already comparatively stable.\\
\begin{figure*}[tb]
 \includegraphics[width=1\linewidth]{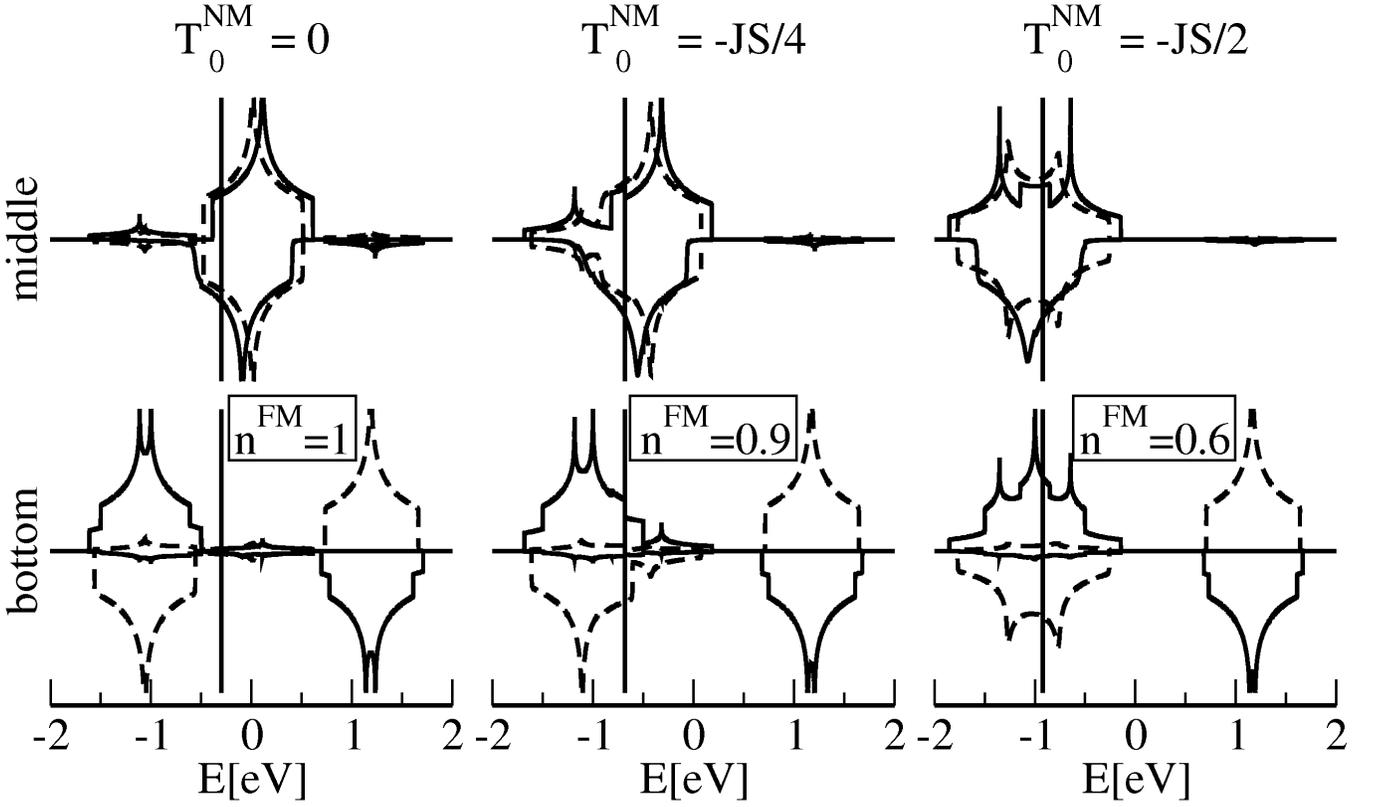}
\caption{\label{figdoslayertnm}Quasi particle density of states of the middle (NM) and bottom (FM) layer for parallel (\emph{solid line}) and antiparallel configuration (\emph{dashed line}) and different positions $T^{\text{NM}}_0$ of the NM layer. Vertical lines mark the chemical potential and the according carrier densities of the FM layers $n^{\text{FM}}$ are shown. Parameters: $S=7.5$, $JS=2$eV, $n=0.8$, $W^{2D}=1$eV, $N_L=3$}
\end{figure*}
Of course it is not required that the FM and NM layers have the same band's center of gravity $T_0$. In fact these two layer types consist in reality of different materials which have another band structure. Thus it is interesting to have different relative positions of the layers. We will focus on the choice $T^{\text{FM}}_0=0$ and an arbitrary $T^{\text{NM}}_0$ in this work.\\
One of the major effects at strong couplings $J$ is the charge transfer from NM to FM layers. This results in a favored occupation of the FM layer and leads to half-filling at $n_C=2/N_L$. This changes when the NM QDOS lies a priori at low energies due to $T^{\text{NM}}_0<0$. The influence of a different position on the QDOS and the charge transfer can be seen in Fig. \ref{figdoslayertnm}. When the NM QDOS comes in the vicinity of the FM QDOS it also gets occupied, leading to a lower density $n^{\text{FM}}$. Especially a half-filling can be removed. Thus even for $n>2/N_L$ and large couplings ferromagnetically ordered phases can appear (cf. Fig. \ref{figpdlayertnmhalb}).
% \begin{figure}[tb]
%  \includegraphics[width=\linewidth]{pdTNMhalb2.eps}
% \caption{\label{figpdlayertnmhalb}Phase diagram as in Fig. \ref{figpdlayer} but with $T^{\text{NM}}_0=-\halb JS$ for $N_L=3$ (\emph{top}) and $4$ (\emph{bottom}) layers. }
% \end{figure}
% \begin{figure}[tb]
%  \includegraphics[width=\linewidth]{exITNMhalb.eps}
% \caption{\label{figexITNMhalb}(color online) Interlayer exchange coupling as in Fig. \ref{figexITNM0} but with $T^{\text{NM}}_0=-\halb JS$ for $N_L=3$ (\emph{top}) and $4$ (\emph{bottom}) layers. Negative values (\emph{red}) mean ferromagnetic and positive (\emph{blue}) antiferromagnetic exchange.}
% \end{figure}
\begin{figure*}
\begin{minipage}{.45\linewidth}
  \includegraphics[width=\linewidth]{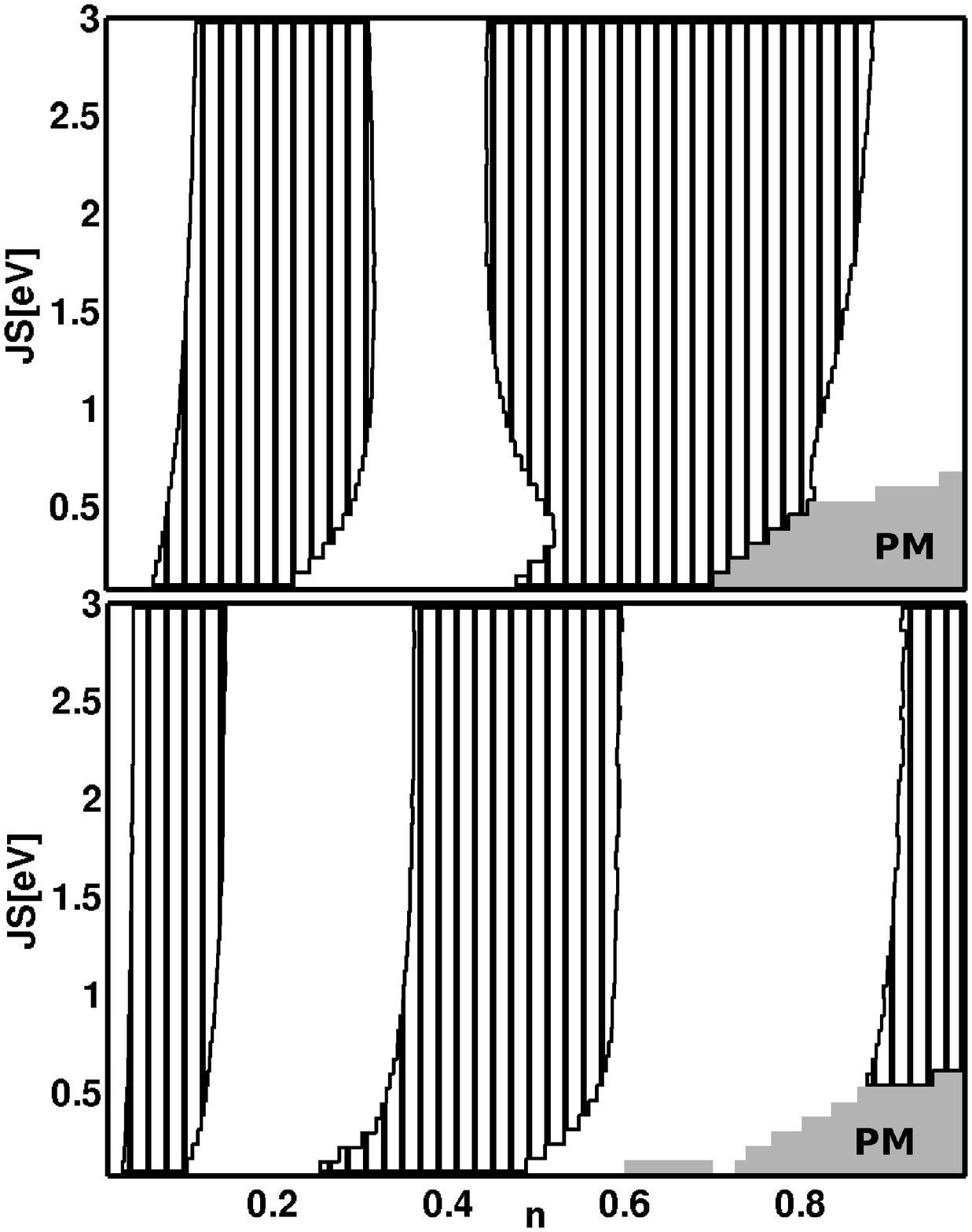}
\end{minipage}
\begin{minipage}{.08\linewidth}
 \ 
\end{minipage}
\begin{minipage}{.45\linewidth}
   \includegraphics[width=\linewidth]{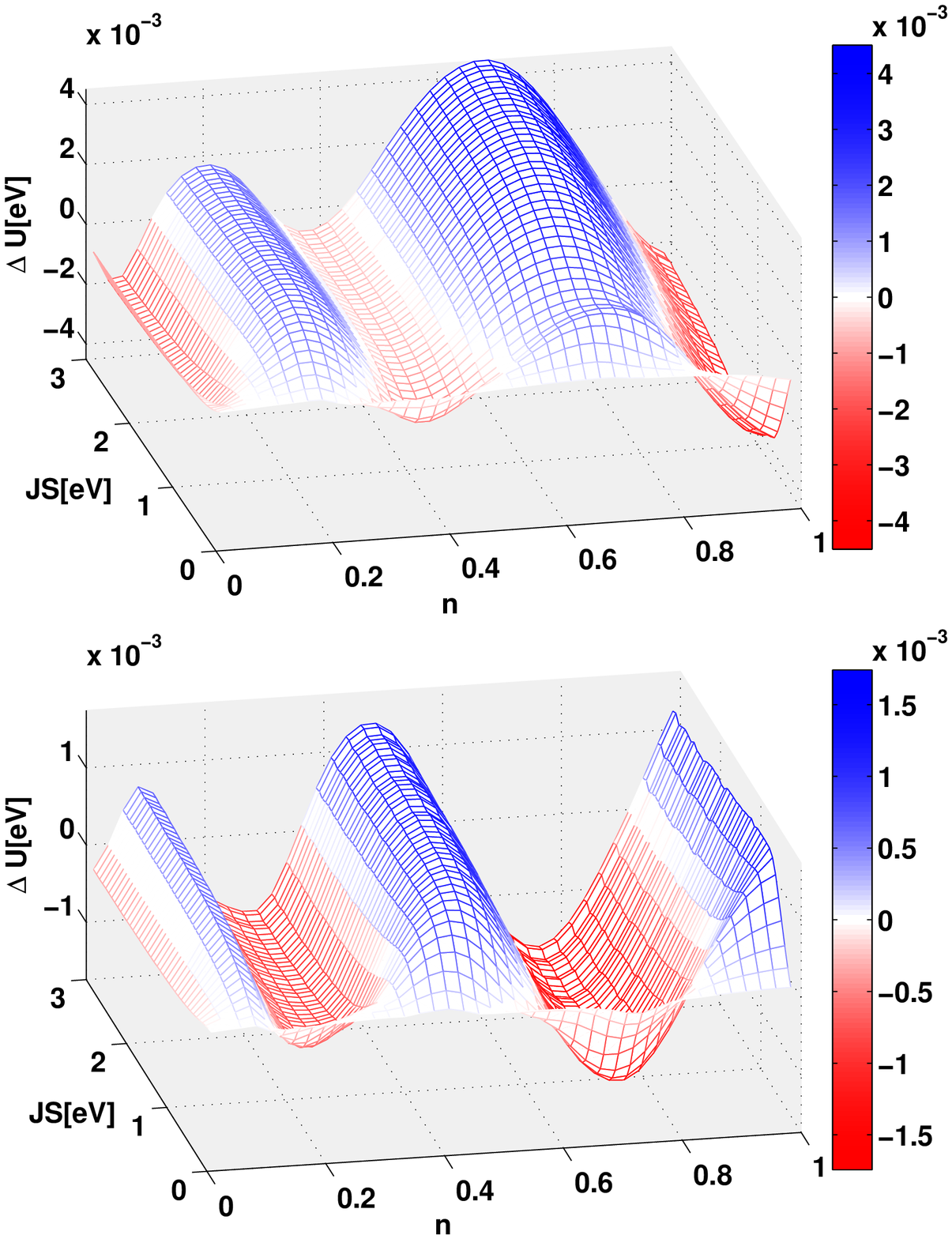}
\end{minipage}
 \caption{\label{figpdlayertnmhalb}(color online) \emph{left}: Phase diagram as in Fig. \ref{figpdlayer} but with $T^{\text{NM}}_0=-\halb JS$ for $N_L=3$ (\emph{top}) and $4$ (\emph{bottom}) layers. \emph{right}: Interlayer exchange coupling as in Fig. \ref{figexITNM0} but with $T^{\text{NM}}_0=-\halb JS$ for $N_L=3$ (\emph{top}) and $4$ (\emph{bottom}) layers. Negative values (\emph{red}) mean ferromagnetic and positive (\emph{blue}) antiferromagnetic exchange.}
\end{figure*}
We use the special choice $T^{\text{NM}}_0=-\halb JS$ which means that the NM QDOS is always at the lower subband of the FM QDOS. Thus the charge transfer does not change at strong couplings any more. Then the system gets in the usual double exchange limit, which means that the phase boundaries do not vary with $J$. However the number of phase areas of parallel and antiparallel alignment still increases with the number of layers $N_L$.\\ 
Not only the charge transfer influences the phase diagram but also the hybridization strength. With increasing coupling and fixed a position of the NM states near $E\approx 0$  the NM/FM layers' QDOSs moved apart from each other and the hybridization was reduced. This especially influenced the magnitude of the interlayer exchange $\Delta U$. When we again use $T^{\text{NM}}_0=-\halb JS$ the NM and FM QDOSs always stay in the same distance and the hybridization strength does not change. So the interlayer exchange is not damped with increasing coupling any more (cf. Fig. \ref{figpdlayertnmhalb}). Instead of that it increases its magnitude at low coupling and saturates at strong couplings. The oscillatory behavior with respect to the electron density is still preserved.

% \begin{figure}
%   \centering
%   \mbox{
%     \subfigure[<subfigure caption here>\label{subfigure label}]{\includegraphics[scale/width]{<filename>}}\quad
%     \subfigure[<subfigure caption here>\label{subfigure label}]{\includegraphics[scale/width]{<filename>}}\quad
%     \subfigure[<subfigure caption here>\label{subfigure label}]{\includegraphics[scale/width]{<filename>}}
%   }
%   \caption{<main figure caption here>}
%   \label{main figure label}
% \end{figure}

\subsection{Curie temperatures}

It is somehow contradictory to calculate transition temperatures in dimension reduced systems, i.e. in $D<3$. The well-known Mermin-Wagner theorem\cite{mermin1966} tells us that  there can be no ordered magnetism in films in the Heisenberg model at finite temperature. This also holds for some other many-body models and especially for the Kondo-lattice model\cite{gelfert2000,gelfert2001}. But the statements are based on the existence of highly symmetric systems, which are not available in reality. Several symmetry-breaking effects (e.g. spin-orbit coupling\cite{pajda2000}) can occur in the experiments or theory which results in finite transition temperatures\cite{sri1998,ney1999,sharma2003ferromagnetism}.  In fact we assume a break of symmetry in the derivation of the entropy $S_0(M)$ where we assume that states with different magnetizations are non-degenerated.  This can be done for example by a (virtual) magnetic field (cf. Sec. \ref{secentropy}). Additionally there can be NM layers in the total system which break the symmetry of the pure KLM, too. Thus the existence of $T_C>0$ in these films is indeed possible.\\
\begin{figure*}[htb]
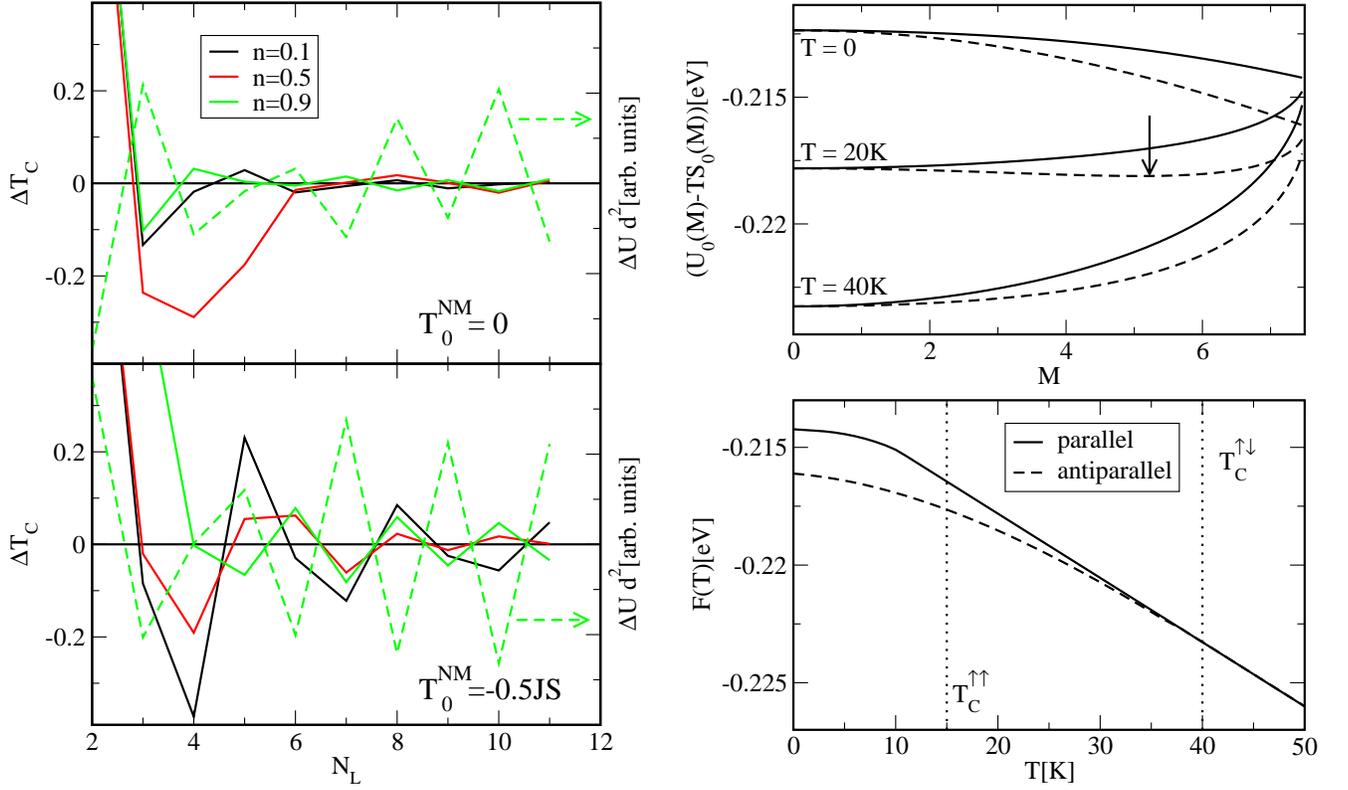

\begin{minipage}{.47\linewidth}
  \includegraphics[width=\linewidth]{deltaTcNL2.eps}
\end{minipage}
\begin{minipage}{.02\linewidth}
 \ 
\end{minipage}
\begin{minipage}{.47\linewidth}
  \includegraphics[width=\linewidth]{umft.eps}
\end{minipage}
\caption{\label{figdeltatc}(color online) \emph{left:} Relative difference of Curie temperatures $\Delta T_C=(T_C^{\uparrow\uparrow}-T_C^{\uparrow\downarrow})/(T_C^{\uparrow\uparrow}+T_C^{\uparrow\downarrow})$ vs. number of layers $N_L$ for $T_0^{\text{NM}}=0$ and $T_0^{\text{NM}}=-\halb JS$. Oscillations are much stronger when the NM layer is near the FM layer (i.e. $T_0^{\text{NM}}=-\halb JS$). Additionally the interlayer exchange $\Delta U d^2$ with the distance $d=N_L-1$ for $n=0.9$ is plotted (dashed line). For $N_L>2$ it shows the typical RKKY-damping ($\Delta U\sim d^{-2}$). \emph{top right:} Reduced (without $I_T(M)$) free energy $\tilde F_T(M)=U_0(M)-TS_0(M)$ for parallel (\emph{solid line}) and antiparallel (\emph{dashed}) alignment at $n=0.2$. The antiparallel alignment is always lower or equal ($M=0$) in energy. The optimal magnetization is at the minimum of $\tilde F_T(M)$ which is for the shown curves at $M^{\text{opt}}=0$ or $M^{\text{opt}}=S$ except for antiparallel alignment at $T=20$K ($M ^{\text{opt}}=5.1$, see arrow). \emph{bottom right}: Resulting free energy at the optimal magnetizations. The alignment with the higher Curie temperature (here $T_C^{\uparrow\downarrow}\approx40$K $>T_C^{\uparrow\uparrow}\approx15$K) is always lower ($T<T_C^{\uparrow\downarrow}$) or equal ($T>T_C^{\uparrow\downarrow}$) in energy. Parameters: $S=7.5$, $W^{2D}=1$eV, $JS=1$eV}
\end{figure*}
It has been seen in the previous section that the parameters $J$ and $n$ have a large influence on the stability of a parallel or antiparallel alignment of the magnetizations in the FM layers to each other. We now want to investigate the behavior at finite temperatures. To do this we have to use the free energy as the thermodynamical potential. In our work it was given as (\ref{eqfm})
\begin{align}
 F_M(T)=&U_M(0)-T(S_M(0)+I_M(T))
\end{align}
which can be mapped on a magnetization dependent function
\begin{align}
  F_T(M)=&U_0(M)-T(S_0(M)+I_T(M))\ . \label{eqmagdepf}
\end{align}
With the condition
\begin{align}
 \frac{\partial F_T(M)}{\partial M}=0
\end{align}
we get
\begin{align}
 T(M)=\left.\frac{\partial_{M'} U_0(M')}{\partial_{M'} S_0(M')+\partial_{M'} I_T(M')}\right|_M\ .\label{eqtm}
\end{align}
This is the temperature when the system has the magnetization $M$. The temperature dependent term $\partial_M I_T(M)$ prevents a direct solution, but can be calculated in principle. It turns out in the Kondo-lattice model that it always holds $\partial_M S_0(M)\gg\partial_M I_T(M)$ and we can neglect $\partial_M I_T(M)$ in a very good approximation. This simplifies (\ref{eqtm}) to
\begin{align}
  T(M)=\left.\frac{\partial_{M'} U_0(M')}{\partial_{M'} S_0(M')}\right|_M
\end{align}
where the right-hand side has no temperature dependence any more. Thus we get a direct equation for the temperature and with it for the magnetization curve $M(T)=T^{-1}(M)$. Especially we get the Curie temperature by choosing
\begin{align}
   T_C=\left.\frac{\partial_{M'} U_0(M')}{\partial_{M'} S_0(M')}\right|_{M=0^+}\ .\label{eqtcdmu}
\end{align}
Let us discuss principal aspects of the formula. At high spins the entropy of the localized moments is much larger than that of the electrons. So we can set $S_0(M)=S_0^{\text{loc}}(M)+S_0^{\text{el}}(M)\approx S_0^{\text{loc}}(M)$. The entropy of the local moments depends only on the magnitude of the magnetization $|M|$ and the spin quantum number $S$. This means that there do not occur any differences between parallel and antiparallel alignment in the entropy. Secondly the differences of the internal energy have to be taken at $M\approx 0$. At $M=0$ there is no difference between the configurations and with it between the energies. It holds
\begin{align}
 \Delta_M U^{\alpha}_0(M)|_{M=0^+}=&U^{\alpha}_0(\Delta M\gtrsim 0) -U^{\alpha}_0(M=0)\\
=&U^{\alpha}_0(\Delta M\gtrsim 0) -U_0(M=0)
\end{align}
 with $\alpha=\uparrow\uparrow,\uparrow\downarrow$. Thus we can calculate the difference of the $T_C$s to
\begin{align}
 \Delta T_C&=T_C^{\uparrow\uparrow}-T_C^{\uparrow\downarrow}\\
&=\frac{U_0^{\uparrow\uparrow}(\Delta M\gtrsim 0)-U_0^{\uparrow\downarrow}(\Delta M\gtrsim 0)}{{\Delta_{M'} S_0(M')}|_{M=0^+}}\ .\label{eqdtc}
\end{align}
The numerator of this fraction is the interlayer exchange $\Delta U$ defined in (\ref{eqdefie}) and $\Delta T_C\sim \Delta U$. The only difference is that we calculated $\Delta U$ in Sec. \ref{sect0} at $M=S$ while in this case we use a small but finite $M=\Delta M\gtrsim0$. The entropy in (\ref{eqdtc})  plays more or less the role of a factor of proportionality.\\
It is well known from RKKY calculations\cite{bruno1992ruderman,slonczewski1995overview} and experiments\cite{ney1999} that the interlayer exchange coupling usually has an oscillatory and damped behavior according to the spacer thickness. The RKKY calculations are based on the Heisenberg model which we do not use in our work. However also in our treatment the oscillations of the interlayer exchange are clearly visible (Fig. \ref{figdeltatc}). In fact the damping goes approximately with the square of the inverse distance $d^{-2}$ as in the RKKY treatment\cite{bruno1991}. Only the direct interlayer exchange between the two FM films ($N_L=2$) is significantly larger than the others. To our opinion the RKKY behavior at $N_L>2$ is recovered because the RKKY is valid for small couplings $J$ and in the PM layers $J$ equals zero. There exists only an indirect interaction due to the hybridization with the FM layers which is small, too. Thus the conditions for an RKKY behavior are fulfilled. In contrast to that a direct coupling of two FM layers at large $J$ is beyond the validity of RKKY calculations. Indeed calculations with modified RKKY integrals show that the interactions range is significantly shorter at strong couplings\cite{carlos2002}.\\
The same oscillations as for $\Delta U$ occur for the (relative) $\Delta T_C=(T_C^{\uparrow\uparrow}-T_C^{\uparrow\downarrow})/(T_C^{\uparrow\uparrow}+T_C^{\uparrow\downarrow})$ (Fig. \ref{figdeltatc}) which is closely related to the interlayer exchange as shown in (\ref{eqdtc}). The connection of the interlayer exchange to an oscillatory behavior of $T_C$ has been found in other work\cite{pajda2000} where the origin of this behavior is traced back to quantum-well states. As for the case of $\Delta U$ at vanishing temperature the differences between $T_C{\uparrow\uparrow}$ and $T_C^{\uparrow\downarrow}$ are heavily damped if the NM layers are at $T_0^{\text{NM}}=0$.\\
When the NM QDOS is near the FM QDOS due to the choice $T_0^{\text{NM}}=-\halb JS$ the amplitudes of the oscillations remain relatively strong even for a large spacer thickness, i.e. a large $N_L$ in our work. The reason is the same as  for $T=0$. Since the QDOSs are near to each other the hybridization is strong and there are large differences between both alignment types. This leads to a large interlayer exchange and with it to a large difference in the Curie temperatures.\\
It has to be said that often both configurations of the magnetizations of the films have a positive transition temperature. Thus both can be the energetically optimal phase at finite temperatures and temperature driven transitions between both alignments could be in principle possible. But usually the phase with the higher transition temperature $T_C^>$ exists in the whole temperature range $[0,T_C^>]$. This can be shown by simple considerations. It has been stated before that there is no difference between the configurations at $M=0$. Thus also the energies of the both types of alignment are equal to each other
\begin{align}
 U_0^{\uparrow\uparrow}(M=0)=U_0^{\uparrow\downarrow}(M=0)\ .
\end{align}
Secondly $U_0(M)$ and also $\partial_MU_0(M)$ are monotonic functions of $M$ (cf. Fig \ref{figdeltatc}) which means that the phase with the lowest internal energy at ferromagnetic saturation $M=S$ also has the lowest energy for all $M$. Due to the fact that the entropy $S_0(M)$ and the integral $I_T(M)$ in (\ref{eqmagdepf})  hardly depend on the type of alignment the free energy is lower for all $M>0$ and $T$ and the transition temperature is higher, too. Thus a phase with higher $T_C$ has a lower free energy for all $M>0$ and is preferred for all $T$ (cf. Fig. \ref{figdeltatc}).\\
% \begin{figure}[htb]
%  \includegraphics[width=\linewidth]{TcJdiffNL.eps}
% \caption{\label{figtcjdiffnl}(color online) Curie temperatures in parallel (\emph{solid lines}) and antiparallel (\emph{dashed}) configuration for $T_0^{\text{NM}}=0$ (\emph{black}) and $T_0^{\text{NM}}=-\halb JS$ (\emph{red}). For $T_0^{\text{NM}}=-\halb JS$ differences between the $T_C$s stay large even for strong couplings $J$.  Parameters: $S=7.5$, $W^{2D}=1$eV, $n=0.3$}
% \end{figure}
% \begin{figure}[htb]
%  \includegraphics[width=\linewidth]{TcndiffNL.eps}
% \caption{\label{figtcndiffnl}(color online) Curie temperatures in parallel (\emph{solid lines}) and antiparallel (\emph{dashed}) configuration for $T_0^{\text{NM}}=0$ (\emph{black}) and $T_0^{\text{NM}}=-\halb JS$ (\emph{red}). For $T_0^{\text{NM}}=-\halb JS$ oscillations are stronger and a finite $T_C$ exists for large $n$ due to suppressed charge transfer.  Parameters: $S=7.5$, $W^{2D}=1$eV, $JS=1$eV}
% \end{figure}
\begin{figure*}[tb]
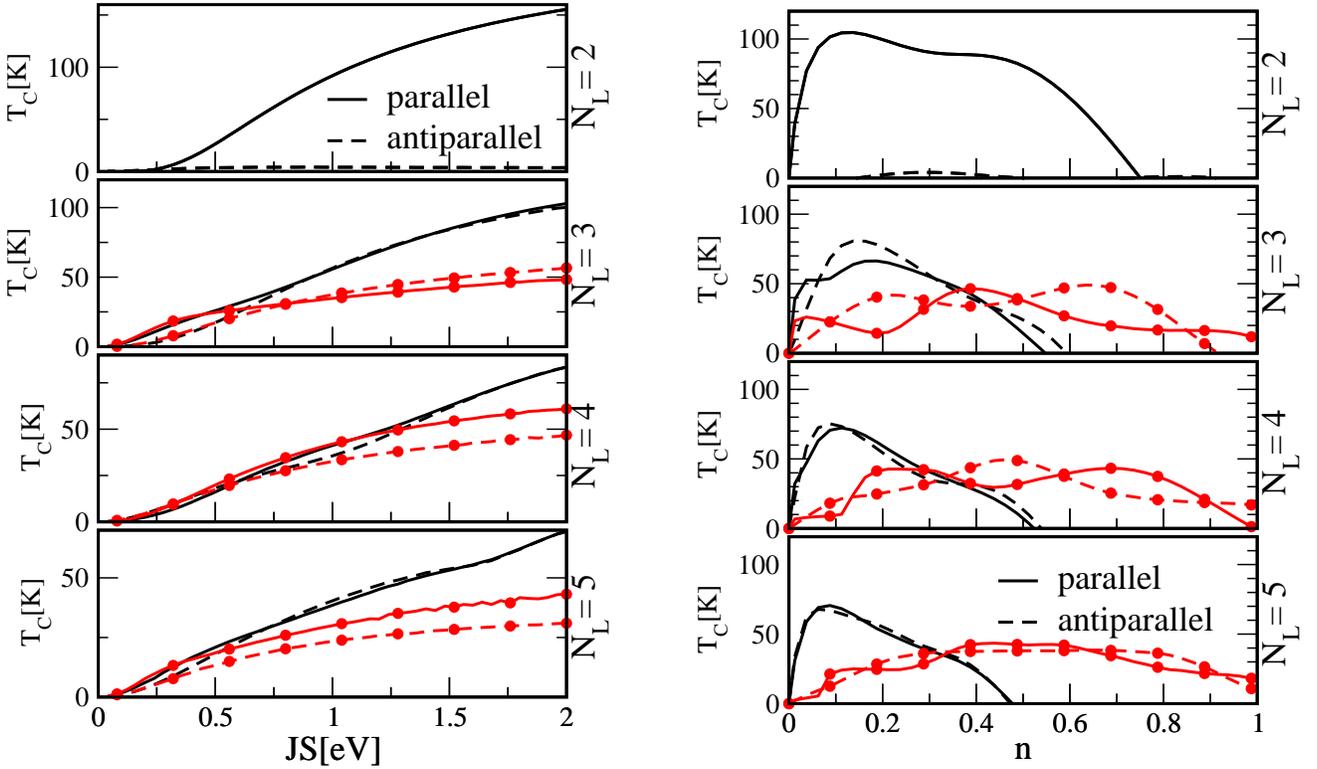

\begin{minipage}{.45\linewidth}
  \includegraphics[width=\linewidth]{TcJdiffNL2.eps}
\end{minipage}
\begin{minipage}{.05\linewidth}
 \ 
\end{minipage}
\begin{minipage}{.45\linewidth}
  \includegraphics[width=\linewidth]{TcndiffNL2.eps}
\end{minipage}
\caption{\label{figtcdiffnl}(color online) Curie temperatures in parallel (\emph{solid lines}) and antiparallel (\emph{dashed}) configuration for $T_0^{\text{NM}}=0$ (\emph{black}) and $T_0^{\text{NM}}=-\halb JS$ (\emph{red}, symbols). \emph{left}: $T_C$ vs. $JS$ at $n=0.3$. For $T_0^{\text{NM}}=-\halb JS$ differences between the $T_C$s stay large even for strong couplings $J$.  \emph{right}: $T_C$ vs. $n$ at $JS=1$eV. For $T_0^{\text{NM}}=-\halb JS$ oscillations are stronger and a finite $T_C$ exists for large $n$ due to suppressed charge transfer.  Parameters: $S=7.5$, $W^{2D}=1$eV}
\end{figure*}
Since the differences of the Curie temperatures are strongly connected to the interlayer exchange, all parameter dependencies are similar to the dependencies of $\Delta U$. The most influential parameters in the Kondo lattice model are the coupling $J$ and the band occupation $n$.\\
Generally in bulk systems $T_C$ increases with $J$ for low and intermediate couplings and saturates at high couplings ("double exchange limit"). This is also true in the film systems treated in this work (cf. Fig. \ref{figtcdiffnl}) whereas the saturation effect is more visible for the case of $T_0^{\text{NM}}=-\halb JS$. At $T_0^{\text{NM}}=0$ the carrier density $n^{\text{FM}}$ in the FM layers gets larger with increasing $J$ due to charge transfer. For those higher densities the saturation tends to start at larger couplings\cite{carlos2002,stier2007}. Again the difference between the Curie temperatures of both configurations is interesting for coupled film systems. We see the typical behavior. For $T_0^{\text{NM}}=0$ the difference decreases even though the absolute values of $T_C$ increase. This is due to the reduction of the hybridization strength because of the larger distance between the FM and NM layer's QDOS. For $T_0^{\text{NM}}=-\halb JS$ the hybridization remains approximately the same and therefore the (absolute) difference between both $T_C$s increases with raising transition temperatures.\\
The role of the charge transfer is more obvious when we look on the influence of $n$ on $T_C$. It can be seen in Fig. \ref{figtcdiffnl} that the range of finite $T_C$ according to $n$ is reduced with $N_L$ when we choose $T_0^{\text{NM}}=0$. It has the same reason as in the case of $T=0$ where only paramagnetism occurred at large couplings and $n>n_C=2/N_L$ due to the half filling of the FM layers (cf. Fig. \ref{figpdlayer}). Since charge transfer to the FM layers is reduced for $T_0^{\text{NM}}=-\halb JS$ or rather there is even transfer to the NM films, the density in the FM layers stays at a moderate level  $n^{\text{FM}}<1$ even at large $n$. Thus also $T_C$ remains positive over the whole range of $n\in [0,1 ]$ for intermediate and strong couplings $J$. Again the maximum differences between $T_C^{\uparrow\uparrow}$ and $T_C^{\uparrow\downarrow}$ remain large at a high layer number $N_L$. As in the case of $T=0$ for the interlayer exchange $\Delta U$ the $T_C$s are strongly oscillating with changing $n$.

\section{Conclusions}

In this work we investigated thin film systems with respect to the alignment of the magnetizations of ferromagnetic layers. We concentrated on "sandwiched" system where two ferromagnetic (FM) layers are separated by an arbitrary number of non-magnetic (NM) layers. The magnetizations of the FM layers could either be parallel or antiparallel to each other. One main focus was on the influence of carriers on the preference of parallel or antiparallel alignment of the magnetizations. Thus we used a Kondo-lattice model to describe interactions of electrons with local moments within the FM layers. With an equation of motion method we could calculate the electrons' properties and with it the internal energy of the system at vanishing temperature $T=0$. By a comparison of the internal energies of the parallel/antiparallel alignments we got magnetic phase diagrams in dependence of the carrier density $n$, the coupling $J$ and the number of layers $N_L$. With increasing number of layers there are more alternations of the stable alignment according to $n$. A very important quantity is the interlayer exchange which we defined as the difference of the internal energies $\Delta U =U^{\uparrow\uparrow}-U^{\uparrow\downarrow}$. Depending on the position of the NM layers $T_0^{\text{NM}}$ the interlayer exchange can either be constant or heavily damped according to an increase of the coupling $J$.\\
Additionally we presented a method to calculate a magnetization dependent free energy $F_T(M)$ without any use of a Heisenberg model. When we minimize $F_T(M)$ according to $M$ we find the optimal magnetization for any given temperature $T$. With it we can calculate Curie temperatures  $T_C^{\uparrow\uparrow}$, $T_C^{\uparrow\downarrow}$ of the respective alignments of the magnetizations. These, or better the differences between them, are strongly connected to the interlayer exchange $\Delta U$. As for $\Delta U$ the differences between the $T_C$s show an oscillating behavior according to $n$. The number of oscillations depend on the number NM layers.\\
The results in this work show, that the carrier density heavily influences the interlayer exchange in the Kondo lattice model, which is maybe not captured in the standard RKKY Heisenberg model treatment. Ab-initio calculations of real materials are in principle exact at $T=0$ and contain more interactions, e.g. superexchange, which makes a direct connection to the pure KLM difficult. But also in ab-initio calculations a mapping onto model systems, e.g. the Heisenberg model, is necessary to get results for $T>0$. This can change the general behavior of the system being the reason why we wanted to avoid such a mapping.\\
It would be very interesting to extend the number of FM layers in the investigated systems or to add semi-infinite leads to the top and bottom layers. This would allow a comparison to bulk systems or potentially lead to new effects. Since these calculations would be numerically extensive they are left for future work.

\appendix
\section{Differences to Mean-field results}

\begin{figure}[tb]
  \includegraphics[width=\linewidth]{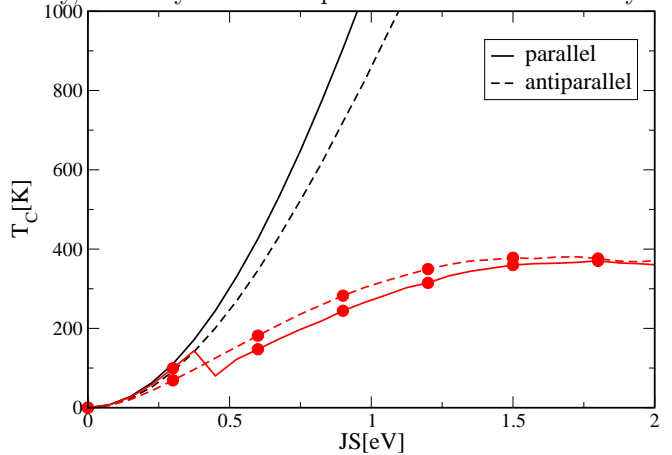}
\caption{\label{figtcjmf}(color online) Curie temperatures in the mean-field approximation for parallel (\emph{solid lines}) and antiparallel (\emph{dashed}) configuration for $T_0^{\text{NM}}=0$ (\emph{black}) and $T_0^{\text{NM}}=-\halb JS$ (\emph{red}, symbols).   The results differ drastically from those of Fig. \ref{figtcdiffnl}. Parameters: $S=7.5$, $W^{2D}=1$eV, $n=0.3$, $N_L=3$}
\end{figure}
The simple mean-field approximation (MFA) of the KLM is often used in literature. In this approximation the spin-flip terms of the Hamiltonian (\ref{eqorigklm}) are neglected resulting in a self-energy
\begin{align}
 M^{\text{MFA}}_{\sigma}(E)=-\zs\halb J M\ .
\end{align}
At $T=0$, or more precisely at ferromagnetic saturation $M=S$, the spin-flip part plays a minor role. Thus the $T=0$ phase diagrams remain almost unchanged in the MFA except that the paramagnetic phase is missing. This has also be found for the pure and diluted KLM\cite{henning2009,dilklm}.\\
When the magnetization is away from saturation the spin-flip part gets important. This is especially true at $T_C$ where the magnetization vanishes and it can be expected that the MFA has several flaws. As an example we want to discuss the $J$-dependence of $T_C$ (cf. Fig. \ref{figtcjmf}). Indeed the results are different from those of Fig. \ref{figtcdiffnl}. For $T^{\text{NM}}=0$ we get the typical MFA dependence $T_C\sim J^2$ which means we get no saturation of $T_C$ at large $J$. Secondly, and maybe more important in terms of interlayer exchange coupling, the difference of the Curie temperatures of different alignments \emph{increases} with increasing $J$. For $T^{\text{NM}}=-\halb JS$ this behavior is reversed - $\Delta T_C$ decreases with increasing $J$. This are completely contrary results compared to the MCDA calculations.\\
The reason for that can be easily understood. Due to the simple structure of the self-energy of the MFA the sub-bands of the QDOS are at $E\approx-\zs\halb JM$ which means their positions depend on the magnetization. Especially for $M=0$ at $T_C$ the sub-bands are at $E\approx 0$. The difference of the $T_C$s depends on the interlayer exchange which is large when the QDOSs of the FM and PM layers are close to each other. In the MFA this is always the case for $M=0$ and $T^{\text{NM}}=0$. Thus $\Delta T_C$ is large for every $J$ and its absolute value increases with the absolute value of $T_C$. For $T^{\text{NM}}=-\halb JS$ the QDOS of the PM layers can be far away from the FM layers' QDOS for a large $J$ and thus $\Delta T_C$ decreases. Because in the MCDA the subbands of the FM layers stay for all $M$ at $E\approx\pm\halb JS$ the situation is inverted - the FM/PM QDOSs are close to each other for $T^{\text{NM}}-\halb JS$ and far away from each other for $T^{\text{NM}}=0$.\\
Thus, besides the drastic overestimation of $T_C$, the mean-field approximation can lead to completely wrong predictions in the KLM\cite{stier3} and should be handled with care. 

\bibliography{films}

\end{document}